\documentclass[11pt,a4paper]{article}
\pdfoutput=1

\usepackage{jcappub} 

\usepackage[T1]{fontenc}
\usepackage{amssymb,amsmath}
\usepackage{bm}

\usepackage{footnote}

\usepackage{bm}

\usepackage{hyperref}

\usepackage{amssymb,amsmath}

\newcommand{\ud}{\,\mathrm{d}}
\newcommand{\uD}{\,\text{D}}

\usepackage{xcolor}

\newcommand{\beq}{\begin{equation}}
\newcommand{\beqn}{\begin{eqnarray}}
\newcommand{\eeq}{\end{equation}}
\newcommand{\eeqn}{\end{eqnarray}}

\usepackage{physics}

\usepackage{multirow}

\newcommand{\be}{\begin{equation}}
\newcommand{\ee}{\end{equation}}

\begin{document}

\title{\center \bf \Huge$\mathcal{N}$-field cosmology in hyperbolic field space: stability and general solutions
}

\author[a]{Perseas Christodoulidis,}
\author[b,c]{Andronikos Paliathanasis}
\affiliation[a]{Van Swinderen Institute for Particle Physics and Gravity, 
University of Groningen, Nijenborgh 4, 9747 AG Groningen, The Netherlands}
\affiliation[b]{Institute of Systems Science, Durban University of Technology, Durban 4000,
South Africa}
\affiliation[c]{Instituto de Ciencias F\'{\i}sicas y Matem\'{a}ticas, Universidad Austral de Chile, Valdivia 5090000, Chile}
\emailAdd{p.christodoulidis@rug.nl}
\emailAdd{anpaliat@phys.uoa.gr}

\abstract{
We study the dynamics of a cosmological model with a perfect fluid and $\mathcal{N}$ fields on a hyperbolic field space interacting via a symmetric potential. We list all late-time solutions, investigate their stability and briefly discuss predictions of the theory. Moreover, for the case of two scalar fields and an exponential potential we prove that the field equations are Liouville integrable and we provide for the first time the general solution for a region of the parameter space.
}


 \begin{flushright}
	98.80.-k, 95.35.+d, 95.36.+x
\end{flushright}
\maketitle

\section{Introduction}

\label{sec1}

Over the last years, two-field models with a symmetric potential and hyperbolic
field space have been extensively studied in the literature in the context of
multi-field inflation \cite{Brown:2017osf,Mizuno:2017idt} or late-time
universe \cite{Cicoli:2020cfj,Cicoli:2020noz}. These models have displayed
interesting phenomenology while remaining observationally viable
\cite{Mizuno:2017idt,Fumagalli:2019noh,Bjorkmo:2019qno,Ferreira:2020qkf,Bounakis:2020xaw}.~\footnote{Here we refer to the predictions of this theory during the inflationary era, whereas the viability of quintessence-like models during late-time cosmology has recently been challenged in the literature (see e.g.~\cite{Basilakos:2019dof,Banerjee:2020xcn}).} Though most works have focused on the two-field regime, certain  many-field constructions have also been proposed as in e.g.~\cite{Bjorkmo:2019aev,Aragam:2020uqi}. Similarly, some progress has been made in the derivation of general solutions
in scalar-field cosmology. On the contrary, multi-field generalizations have been proved more challenging and up to date only a few solutions are known for arbitrary number of fields \cite{Chimento:1998ju,Christodoulidis:2018msl,Paliathanasis:2018vru,Socorro:2020nsm,Socorro:2021bco}.

The existence of exact and analytic solutions is an essential property for the
mathematical description of a physical theory. Although a dynamical system can
be solved by using numerical techniques we do not know that the numerical
trajectories correspond always to real solution of the problem, thus we should investigate if the dynamical system posses the integrability property. There are various techniques for the study of the integrability in the literature. In cosmological studies,
due to the fact that the gravitational field equations for scalar field
theories admit a minisuperspace description, techniques from analytic mechanics
can be applied. The theory of similarity transformations for the derivation of
conservation laws has been applied in \cite{ns1,ns2,ns3,ns4,ns5} while some
other approaches can be found in \cite{ns6,ns7,ns8}. Another important
approach for the study of a cosmological model is the determination of the
stationary points. The latter points can be used for the determination of the
asymptotic behaviour of a specific theory and to extract important information
and criteria for the cosmological evolution of the specific model
\cite{dn1,dn2,dn3,dn4,quin00}.

In this work we will investigate the $\mathcal{N}$-field generalization of the
two-field hyperbolic problem in the presence of a perfect fluid. We will first list
all critical-point solutions and investigate their stability. Next, we will
apply the Noether method to derive two-field general solutions in some cases
and then extend them to $\mathcal{N}$-fields.

The paper is organized as follows: in Sec.~\ref{sec:two_fields} we revisit the
stability analysis for the two-field hyperbolic problem in the presence of a
fluid. Next, we generalize the discussion for $\mathcal{N}$ fields in
Sec.~\ref{sec:n_field} and list all new solutions as well as their stability
properties. In Sec.~\ref{sec:general_solution} we calculate the general
solution for a subset of the parameter space using the Noether method.
Finally, in Sec.~\ref{sec:conclusions} we offer our conclusions.

\section{Chiral (multi-field) cosmology}

The Chiral cosmological model belongs to the family of the Einstein-nonlinear
$\sigma$-model where in the Einstein-Hilbert action two scalar fields
minimally coupled to gravity are introduced such that the gravitational action
integral can be written as follows \cite{chir3,sigm0,sigm1,ch1}%
\begin{equation}
S=\int\sqrt{-g} \ud^4 X\left(  R-\frac{1}{2}g^{\mu\nu}\nabla_{\mu}\phi\nabla_{\nu}%
\phi-\frac{1}{2}g^{\mu\nu}F\left(  \phi\right)  \nabla_{\mu}%
\psi\nabla_{\nu}\psi-V\left(  \phi\right)  \right)  +S_{m} \, , \label{acc.01}%
\end{equation}
where there is a coupling in the kinetic term between the two scalar fields.
When the coupling function $F\left(  \phi\right)  $ is constant the Action
Integral (\ref{acc.01}) describes two quintessence fields; however, in
Chiral cosmology the two dynamics of the two fields evolve in a space of
constant non-zero curvature, that is, $F\left(  \phi\right)  =e^{\kappa\phi}$.
At this point it is important to mention that we refer to a two-dimensional
space defined by the kinetic terms of the scalar fields and not in the
background space with metric $g_{\mu\nu}$ and Ricci scalar $R$.

According to the cosmological principle for the background space we assume
that of spatially flat Friedmann--Lema\^{\i}tre--Robertson--Walker (FLRW)
described by the line element
\begin{equation}
\ud s^{2}=-N_l\left(  t\right)  ^{2} \ud t^{2}+a\left(  t\right)  ^{2}\left(
 \ud X^{2}+ \ud Y^{2}+ \ud Z^{2}\right)  ,
\end{equation}
where $a\left(  t\right)  $ is the scale factor and $N_l\left(  t\right)  $ the
lapse function. Moreover, we assume that the scalar fields inherit the
symmetries of the background space, that is, $\phi\left(  x^{\mu}\right)
=\phi\left(  t\right)  $, $\psi\left(  x^{\mu}\right)  =\psi\left(  t\right)
$, while the Action Integral $S_{m}$ describes an ideal gas with energy
density $\rho$, pressure $p$ and constant equation of state parameter
$p=w\rho$.

Hence, the gravitational field equations are \cite{ch1}%
\begin{align}
3H^{2}  & =\rho_{f}+\rho,\\
-\left(  2\dot{H}+3H^{2}\right)    & =p_{f}+P,
\end{align}
in which $\rho_{f},$ $p_{f}$ are the energy density and pressure components of
the two scalar fields, that is,\qquad\
\begin{align}
\rho_{f}  & =\frac{1}{2}\dot{\phi}^{2}+\frac{1}{2}e^{\kappa\phi}%
\dot{\psi}^{2}+V\left(  \phi\right)  ,\\
p_{f}  & =\frac{1}{2}\dot{\phi}^{2}+\frac{1}{2}e^{\kappa\phi}%
\dot{\psi}^{2}-V\left(  \phi\right)  .
\end{align}
Finally, the equations of motion for the two fields are
\begin{align}
\ddot{\phi}+3H\dot{\phi} - \kappa\frac{1}{2}%
e^{\kappa\phi}\dot{\psi}^{2}+V_{,\phi}\left(  \phi\right)  =0 \, , \\
\ddot{\psi}+3H\dot{\psi}+\kappa\dot{\phi}\dot{\psi}=0~,\label{sp.07}%
\end{align}
while the ideal gas satisfies the continuity equation%
\begin{equation}
\dot{\rho}+3H(1+w)\rho=0\,,
\end{equation}
from which it follows $\rho=\rho_{m0}a^{-3\left(  1+w\right)  }$.

At this point it is important to mention that the gravitational field
equations can be derived by the variation of the point-like Lagrangian
\begin{equation}
L\left(  a,\dot{a},\phi,\dot{\phi},\psi,\dot{\psi}\right)  =\frac{1}{2N_l \left(
t\right)  }\left(  -6a\dot{a}^{2}+a^{3}\left(  \dot{\phi}^{2}+
e^{\kappa\phi}\dot{\psi}^{2}\right)  \right)  -N\left(  t\right)
a^{3}V\left(  \phi\right)  +N_l \left(  t\right)  \rho_{m0}a^{-3w_{m}%
}.\label{ssl1}%
\end{equation}

That it is an important observation, because we can apply techniques from
analytic mechanics for the determination of exact solutions. As far as the
scalar field potential $V\left(  \phi\right)  $ is concerned, we will assume the
exponential function $V\left(  \phi\right)  =V_{0}e^{\lambda\phi}$ for the most part of this paper which has been shown to provide interesting physical results \cite{ancqg,Christodoulidis:2019jsx}.

In order to study the asymptotic behaviour of the previous model it is better to switch from cosmic time to the e-folding number defined from $\ud N = H \ud t$. In this way the set of evolution equations for the scalar fields and the fluid can be written as an autonomous dynamical system. Introducing a new variable $z=\rho/H^2$, describing the fluid energy density,
the Friedman constraint becomes
\begin{equation}
3 = {1 \over 2} v^iv_i  + z + V\, ,
\end{equation}
which implies that the allowed values for the field velocities and $z$ should satisfy
\begin{equation}
 {1 \over 2} v^iv_i  + z \leq 3 \, .
\end{equation}
The slow-roll parameter becomes
\begin{equation}
\epsilon = {1 \over 2}v^iv_i + {1\over 2}(1+w)z \, ,
\end{equation}
while the potential satisfies
\begin{equation}
{V \over H^2} = 3 - \epsilon + {1\over 2}(w-1) z \, .
\end{equation}
The evolution equations for the fields and the fluid are
\begin{align} \label{eq:dvi}
&(v^i)' + \Gamma^i_{jk}v^jv^k + (3 - \epsilon)(v^i + \lambda^i) + {1\over 2}(w-1) \lambda^i z =0\, , \\
&z' + (3 + 3w -  2\epsilon)z =0 \, .
\end{align}
However, critical points for the velocities are not expected to be found when a generic field metric is considered because Christoffel symbols are field dependent. Instead, it is better to study equations for the normalized velocities $\sqrt{G_{ii}} v^i$ (no sum is assumed in $i$) that enter the definition of $\epsilon$. For the field metric with an isometry these velocities are defined as
\begin{equation}
y \equiv \phi' \, , \qquad x \equiv \sqrt{F(\phi)}\psi' \, .
\end{equation}
The variables of the two-field dynamical system in first order form are $\phi,\psi,y,x,z$.

\section{Recap of the two-field problem with a perfect fluid} \label{sec:two_fields}

\subsection{Hyperbolic field metric}

Specializing to the hyperbolic field metric
\begin{equation}
\ud s^2 = \ud \phi^2 + e^{\kappa \phi} \ud \psi^2 \, ,
\end{equation}
the system in first order form is
\begin{align} \label{eq:dphi_2f}
&\phi' = y \, , \\
&\psi' = x e^{-\kappa/2 \phi} \, , \\ \label{eq:dy_fluid}
&y' + (3-\epsilon)(y+\lambda) - {\kappa \over 2} x^2 + {1\over 2}(w -1) \lambda z =0\, , \\  \label{eq:dx_fluid}
&x' + \left( 3 - \epsilon + {\kappa \over 2}y \right) x = 0 \, , \\ \label{eq:dz_fluid}
&z' + (3 + 3w - 2\epsilon)z =0 \, .
\end{align}
The second equation can be discarded because $\psi$ is a cyclic variable and does not affect dynamics and, moreover, we observe that the last three equations do not depend on $\phi$ and so the first equation can also be omitted; the reduced $3 \times 3$ system is sufficient to extract information regarding the solution. Although the stability analysis for this model has been presented recently in Ref.~\cite{Cicoli:2020cfj}, in this section we will revisit it in order to facilitate the transition to more fields. In addition, we will mention the stability criteria for generic $w$ that were missing for some solutions of the aforementioned work.

For our set of variables the eigenvalues of the Jacobian matrix follow straightforwardly from our analysis because that matrix can always be written in block-diagonal form. We have the following critical points:

\begin{enumerate}
\item First, we have scalar field domination solutions, which generalize the solutions presented in \cite{Christodoulidis:2019jsx,Dimakis:2019qfs} with the addition of $z=0$. Their stability properties remain unchanged with the additional requirement $\epsilon <3/2(1+w)$. This happens because the stability matrix acquires an upper triagonal form with zeros below the main diagonal and so the presence of the fluid does not affect eigenvalues (for this type of solutions). There are three types of scalar-dominated solutions:

\begin{enumerate}
\item The \textit{scalar-field gradient} solution,
\begin{equation}
(y,x,z)_{\rm gr} = \left(-\lambda, 0 ,0\right) \, .
\end{equation}
Motion is aligned with the potential gradient flow. It is stable provided
\begin{equation}
-\sqrt{6 + {\kappa^2 \over 4}} - {\kappa \over 2} < \lambda < \sqrt{6 + {\kappa^2 \over 4}} - {\kappa \over 2} \, , \qquad |\lambda| < \sqrt{6} \, ,  \qquad |\lambda| < \sqrt{3(1+w)}\, .
\end{equation}
The solution is depicted at the left panel of Fig.~\ref{fig:plota}.
\begin{figure}[t!]
\centering
\includegraphics[width=.45\textwidth]{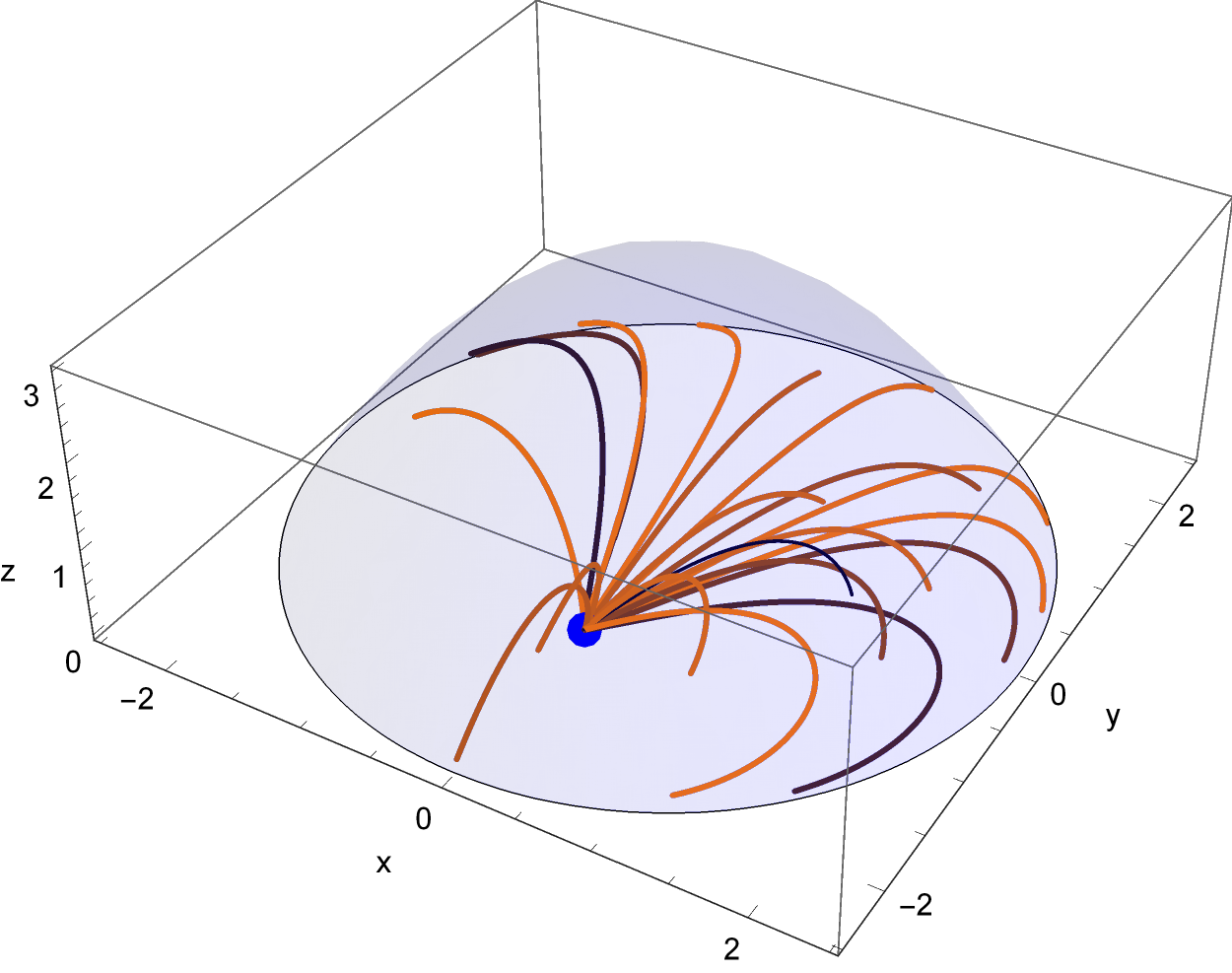}
\includegraphics[width=.45\textwidth]{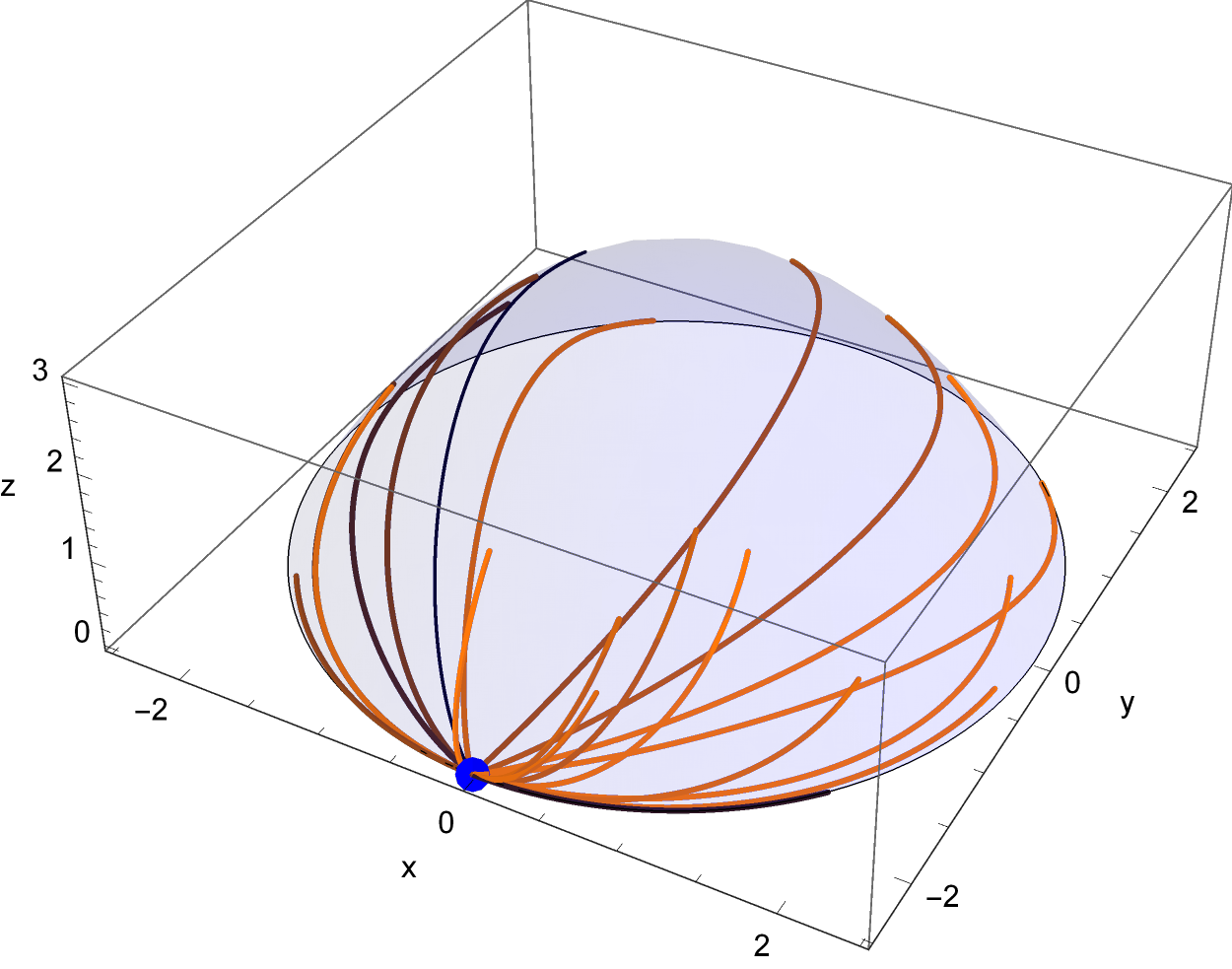}
\caption{ \it The numerical solutions for a wide range of initial conditions (drawn uniformly from the surface defined from $\epsilon\approx3$). Left: For $\lambda=\kappa=1$ and $w=0$ the solution asymptotes to the gradient critical point. Right:  For a fluid with $w=2$, $\lambda=3$ and $\kappa=-1$ the solution asymptotes to kinetic domination. Blue dots correspond to the respective critical points and the semi-transparent blue surface denotes the region of definition for $x,y,z$.}
\label{fig:plota}
\end{figure}

\item The \textit{scalar-field kinetic domination} solution given as
\begin{equation}
(y,x,z)_{\rm kin} = \left(\pm \sqrt{6}, 0 ,0\right) \, ,
\end{equation}
and is stable for
\begin{equation}
|\lambda|>\sqrt{6}\, , \qquad \lambda \cdot \kappa<0\, , \qquad 1+w>2\, ,
\end{equation}
(see left panel of Fig.~\ref{fig:plota}).

\item The \textit{hyperbolic} solution
\begin{equation}
(y,x,z)_{\rm hyper}  = \left( - {6 \over \kappa + \lambda}, \pm{\sqrt{6} \sqrt{\lambda^2 + \kappa \lambda   -6 }  \over \kappa + \lambda} ,0 \right)  \, .
\end{equation}
This solution exists provided $k\neq-\lambda$, $\lambda^2 + \kappa \lambda   -6>0$ or
\begin{equation}
\lambda>\sqrt{6 + {\kappa^2 \over 4}} - {\kappa \over 2} \, , \qquad \lambda < -\sqrt{6 + {\kappa^2 \over 4}} - {\kappa \over 2} \, ,
\end{equation}
and it is stable whenever the previous two solutions are unstable, namely when the following conditions are satisfied
\begin{align} \label{eq:hyper1}
&\kappa>0\, , \qquad \lambda>\sqrt{6 + {\kappa^2 \over 4}} - {\kappa \over 2} \, , \qquad {\lambda \over \kappa +  \lambda}<{1 \over 2} (1+w) \\ \label{eq:hyper2}
& \kappa<0 \, , \qquad \lambda < -\sqrt{6 + {\kappa^2 \over 4}} - {\kappa \over 2} \qquad {\lambda \over \kappa +  \lambda}<{1 \over 2} (1+w) \, .
\end{align}
The solution is depicted at the left panel of Fig.~\ref{fig:plotb}.
\end{enumerate}
\begin{figure}[t!]
\centering
\includegraphics[width=.49\textwidth]{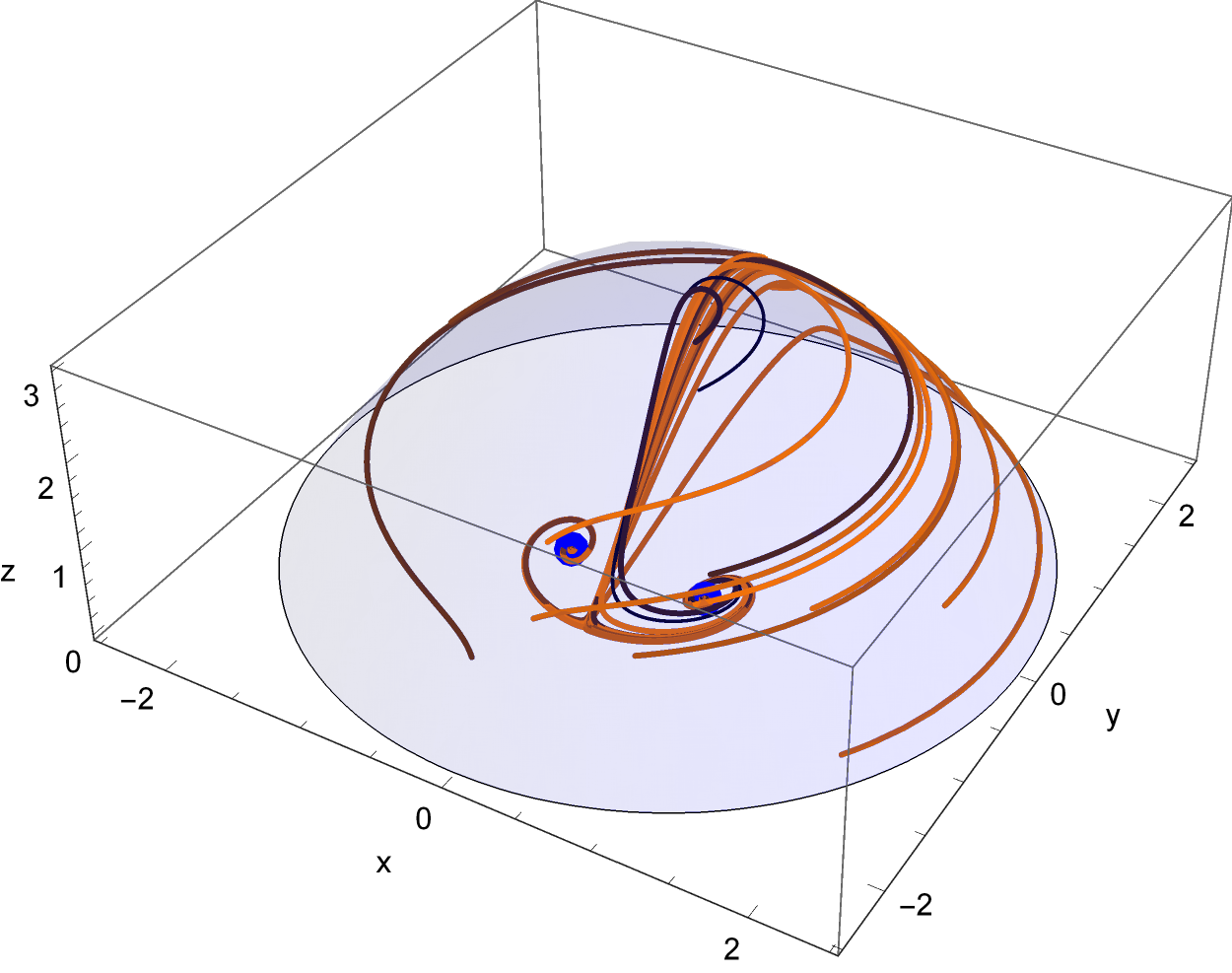}
\includegraphics[width=.49\textwidth]{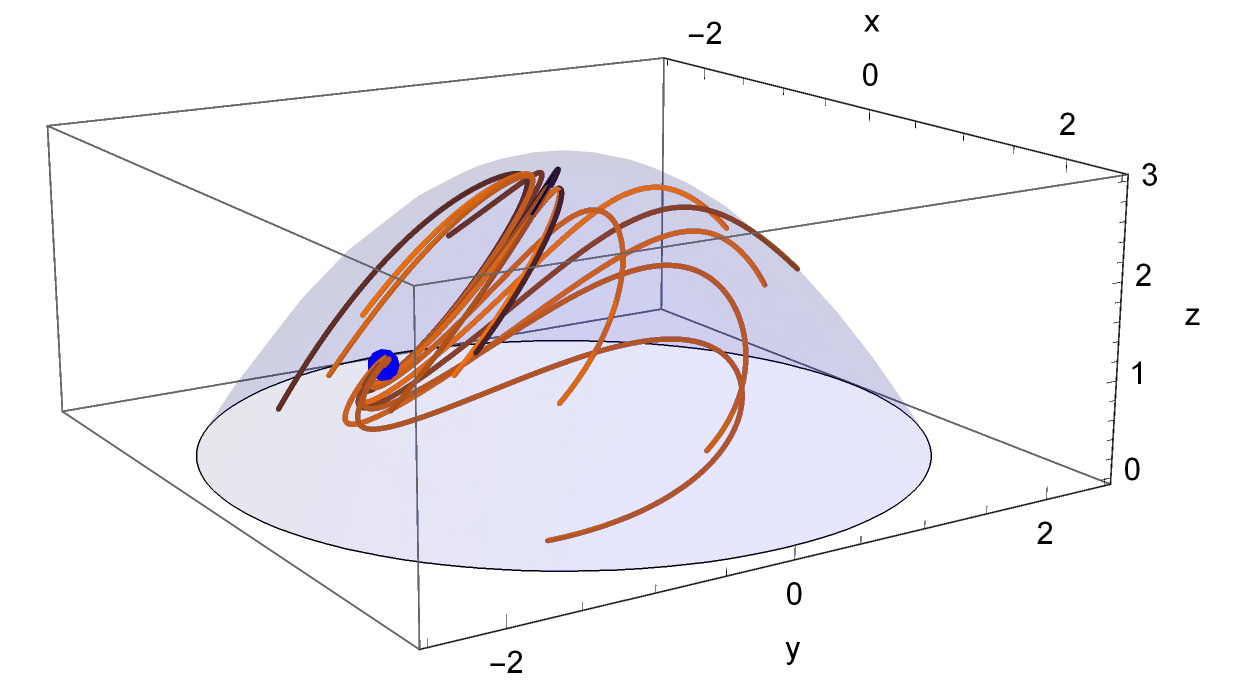}
\caption{ \it
The numerical solutions assuming a fluid with $w=0$ for a wide range of initial conditions (drawn uniformly from the surface defined from $\epsilon\approx3$). Left: For $\lambda=1$, $\kappa=15$ the solution asymptotes to one of the two hyperbolic critical points (depending on the initial $x$). Right:  For $\lambda=2.2$, $\kappa=1$ the solution asymptotes to the scaling critical point. Blue dots correspond to the respective critical points and the semi-transparent blue surface denotes the region of definition for $x,y,z$.}
\label{fig:plotb}
\end{figure}

\item The second type of solutions describe fluid domination $z=3$, where fields have zero values. In order for this to be a solution of the dynamical system the parenthesis of Eq.~\eqref{eq:dz_fluid} should be zero and this gives the value of $\epsilon$
\begin{equation}
\epsilon = {3 \over 2} (1+w)\, ,
\end{equation}
(this value of $\epsilon$ is compatible with Eq.~\eqref{eq:dy_fluid}). These solutions are stable for $w<-1$, which yields $\epsilon<0$, and thus describe contracting universes.

\item Finally, we find the scaling solution with the fluid and at least one of the fields non-zero. Again we require the parenthesis of Eq.~\eqref{eq:dz_fluid} to vanish and so $\epsilon$ has the same value as in fluid domination. The solution is
\begin{equation}
(y,x,z)_{\rm scal} = \left( -{3 (w+1)  \over \lambda},  0 ,3 - {9(1+w) \over \lambda^2} \right) \, ,
\end{equation}
which exists provided $z_{\rm scal}\geq 0$ or
\begin{equation}
\lambda^2 \geq 3(1+w) \, .
\end{equation}
The eigevalues of the Jacobian matrix are
\begin{align}
m_1 &= {3 \over 2\lambda} \left[ (w+1)\kappa +(w-1)\lambda \right] \, ,\\
m_{\pm} &= {3 \over 4} \left(w-1 \pm \sqrt{(w-1)^2 +8(w^2-1) -24 (w-1)(w+1)^2 \lambda^{-2}} \right) \, ,
\end{align}
and they are non-positive for $w>-1$ with the following additional restrictions on $w,\lambda,\kappa$
\begin{align}
-1<w<1\, , \qquad  {\kappa \over \lambda} < {1 - w \over 1+w}  \, .
\end{align}
For $w=0$ we recover the relations mentioned in Ref.~\cite{Cicoli:2020cfj}. The solution is illustrated at the right panel of Fig.~\ref{fig:plotb}. Note that there are no real solutions with $y,x,z\neq0$.

\end{enumerate}

\subsection{Generic field metric with isometry} \label{subsec:generic_2_field_metric}
\begin{figure}[t!]
\centering
\includegraphics[width=.49\textwidth]{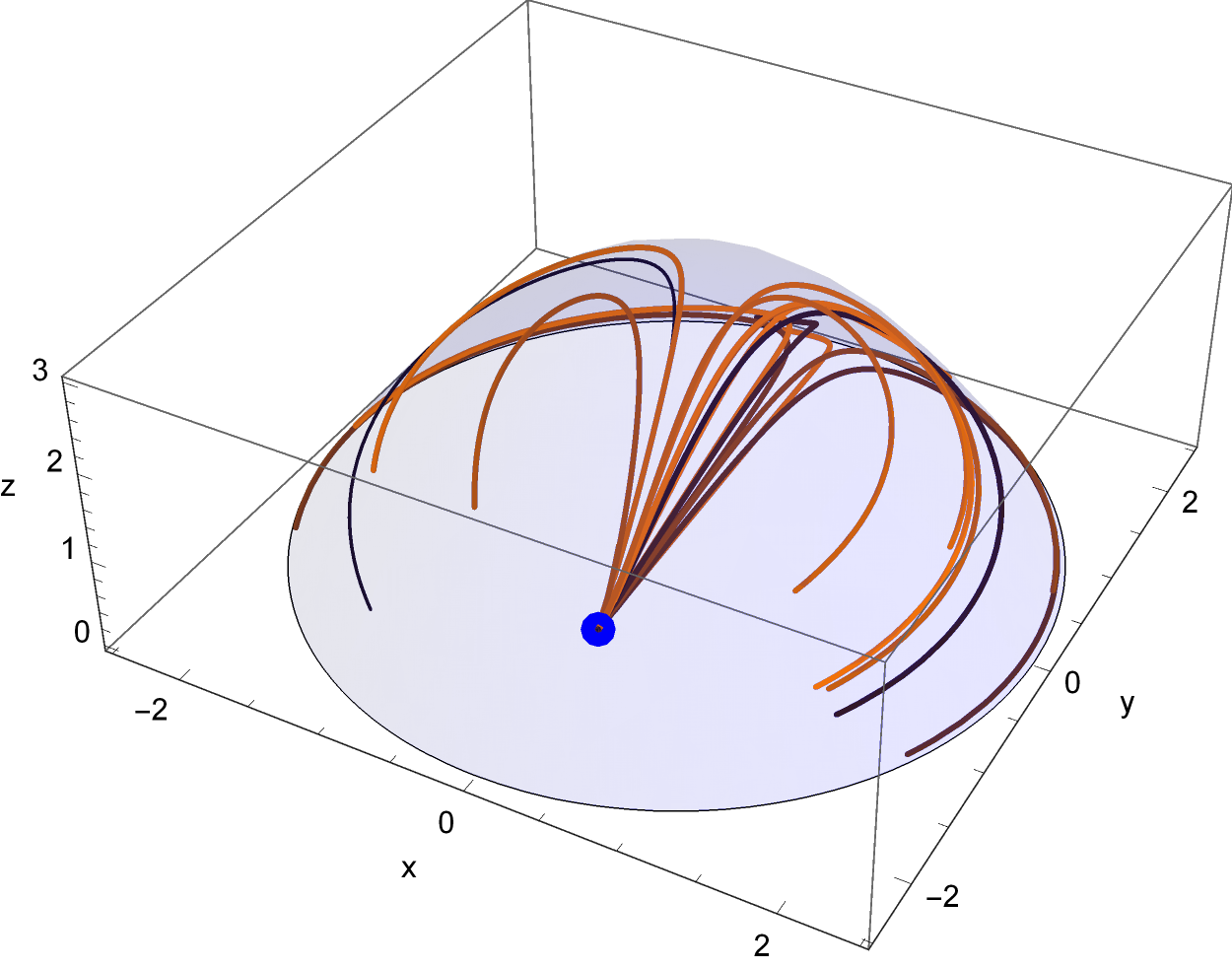}
\includegraphics[width=.49\textwidth]{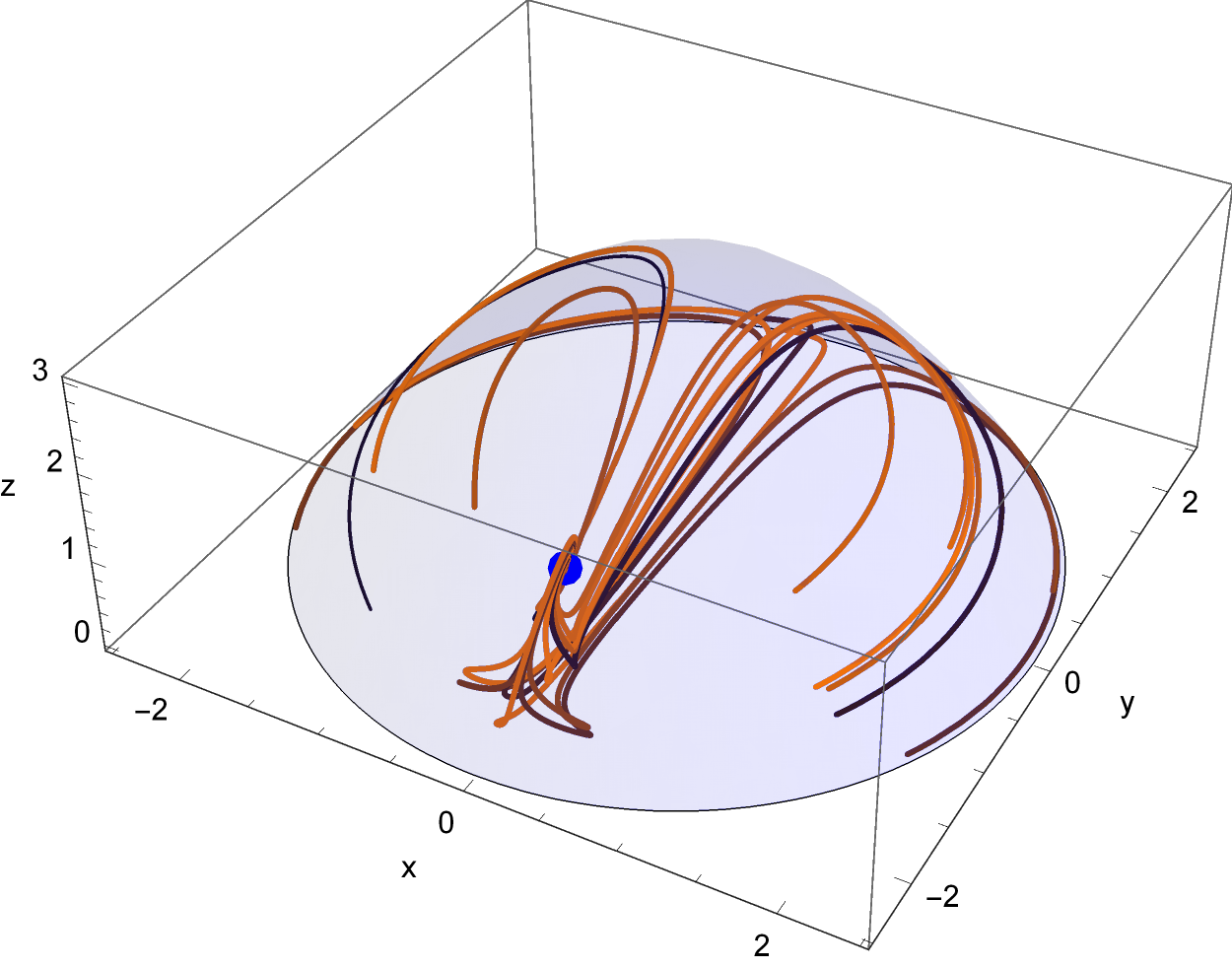}
\caption{ \it
The numerical solutions assuming a fluid with $w=0$ and $\phi_0=2$ for a wide range of initial conditions (drawn uniformly from the surface defined from $\epsilon\approx3$). Left: For $\lambda=1$ the solution asymptotes to the gradient critical point. Right:  For $\lambda=2.2$  the solution asymptotes to the scaling critical point. Blue dots correspond to the respective critical points and the semi-transparent blue surface denotes the region of definition for $x,y,z$.
}
\label{fig:plot_g_m}
\end{figure}

\begin{figure}[t!]
\centering
\includegraphics[width=.49\textwidth]{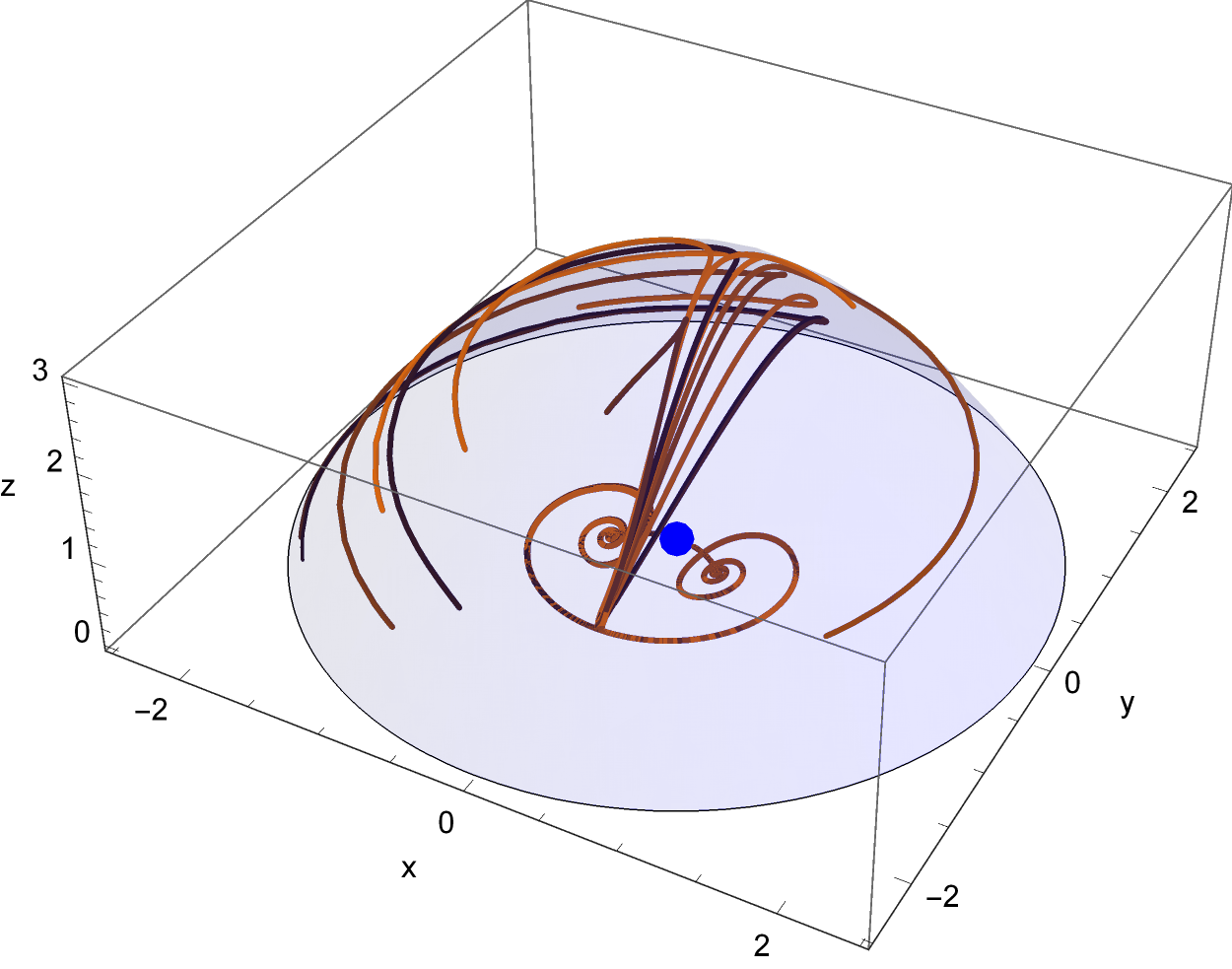}
\caption{ \it
Numerical solutions that asymptote to the de Sitter critical point with $\phi_0=2$, $\lambda=1$.}
\label{fig:plot_g_m2}
\end{figure}

We will briefly comment on the case of a general field metric with isometry in $\psi$
\begin{equation}
\ud s^2 = \ud \phi^2 + F(\phi)\ud \psi^2 \, .
\end{equation}
In this case $\kappa \equiv F_{,\phi}/F$ is field dependent and Eq.~\eqref{eq:dphi_2f} can not be omitted. Choosing an exponential potential, or a potential that asymptotes to an exponential at e.g.~$-\infty$, all previous solutions (except for the hyperbolic one) may exist for appropriate choices of the metric function (see Fig.~\ref{fig:plot_g_m} for a model with $F=e^{\phi^2}$), while the hyperbolic solution is replaced by a de Sitter asymptotic state $y,x\rightarrow0$.
This can be shown as follows: if the gradient or kinetic solutions are unstable then a solution, that resembles the hyperbolic one, can be obtained if $\kappa$ diverges to plus/minus infinity at the boundary of the space. In this case the combination $\kappa y$ is required to be constant and so the parenthesis of Eq.~\eqref{eq:dx_fluid} can vanish. Plugging back into Eq.~\eqref{eq:dy_fluid} gives the asymptotic solution for $y$ and $x$ which has exactly the same form as in the hyperbolic case, albeit $\kappa$ is field dependent and growing in norm \cite{Christodoulidis:2019jsx,Cicoli:2020noz}. Even though a proper stability analysis for a field-dependent $\kappa$ requires study of the dynamical system at infinity, we can use a simpler argument to understand the behaviour of these solutions. For the $4\times4$ dynamical system calculating Lyapunov exponents, results to one zero eigenvalues which is associated with $\phi$. Since $\phi$ will eventually roll towards decreasing values of the potential (otherwise the Friedman constraint would be violated) no instability related to this marginal direction will be present. Therefore, one can apply the previous formulae for solutions and stability criteria after taking the limit $\phi \rightarrow \infty$. Note though that convergence towards the de Sitter critical point is much slower compared to other critical points, while the system may pass through other critical points first (see Fig.~\ref{fig:plot_g_m2} for a model with $F=e^{\phi^3}$ as well as the discussion in Ref.~\cite{Cicoli:2020noz}).

\section{The $\mathcal{N}$- field hyperbolic solution} \label{sec:n_field}
To properly generalize the model to $\mathcal{N}$ fields we choose the following form of the field metric
\begin{equation}
\ud s^2 = \ud \phi^2 +\sum e^{\kappa_i \phi} \ud \psi_i^2 \, ,
\end{equation}
with Ricci scalar
\begin{equation}
R = -{1 \over 4} \left(\sum \kappa_i \right)^2 - {1 \over 2} \sum \kappa_i^2 \, .
\end{equation}
Equations of motion are
\begin{align} \label{eq:dphi}
&\phi' = y \, , \\ \label{eq:chii}
&\psi'_i = x_i e^{-\kappa_i \phi} \, , \\ \label{eq:dy_fluid2}
&y' + (3-\epsilon)(y+\lambda) - \sum{\kappa_i \over 2} x_i^2 + {1\over 2}(w -1) \lambda z =0\, , \\ \label{eq:dxi}
&x_i' + \left(3 - \epsilon + {\kappa_i \over 2}y \right) x_i = 0 \, , \\ \label{eq:dz_fluid2}
&z' + (3 + 3w - 2\epsilon)z =0 \, .
\end{align}
To analyse the problem we will distinguish between two cases.

\subsection{$\kappa_i$ are all equal}
In the symmetric case $\kappa_i= \kappa$ the solutions presented in Sec.~\ref{sec:two_fields} carry over with the substitution $x^2\rightarrow \sum x_i^2$. This becomes apparent if we write the differential equation of the slow-roll parameter $\epsilon$ which is found by contracting Eq.~\eqref{eq:dvi} with $v_i$:
\begin{equation} \label{eq:depsilon}
\epsilon' +(3-\epsilon)(2\epsilon + \lambda y) + {1\over 2}(w -1) \lambda y z =0 \, ,
\end{equation}
and substituting $\sum x_i^2 = 2\epsilon -y^2$ in Eq.~\eqref{eq:dy_fluid2}
\begin{equation}
y' + (3-\epsilon)(y+\lambda) - {\kappa \over 2}(2\epsilon -y^2) + {1\over 2}(w -1) \lambda z =0 \, .
\end{equation}
This shows that the set of Eqs.~\eqref{eq:dxi} can be replaced with Eq.~\eqref{eq:depsilon} and $x_i$ are left undetermined (a similar argument was used in Ref.~\cite{Paliathanasis:2020abu}).

\subsection{$\kappa_i$ are different}
The situation is drastically different when $\kappa_i$ are different. In the next we list all critical-point solutions and their stability properties.

\begin{enumerate}
\item The analogue of the hyperbolic solution with $y$ and all $x_i$ different than zero is inconsistent as it requires
\begin{equation}
(3 - \epsilon) + {\kappa_i \over 2}y = 0 \, ,
\end{equation}
to hold for every $\kappa_i$. Therefore, we conclude that only one $x_j$ can be non-zero and $x_i=0$ for $i\neq j$. To study the stability we calculate the Jacobian matrix evaluated on the hyperbolic solution and we observe that it always acquires a block diagonal form (some permutations of rows and columns may be necessary)
\begin{equation}
\begin{pmatrix}
A_{2\times2} & 0_{2\times \mathcal{N}-1} \\
0_{\mathcal{N}-1 \times 2}  & B_{\mathcal{N}-1 \times \mathcal{N}-1}
\end{pmatrix} \, ,
\end{equation}
where
\begin{equation}
A_{2\times2} =  {1 \over (\kappa_j + \lambda)^2}
\begin{pmatrix}
36 - 3(\kappa_j-\lambda)(\kappa_j+2\lambda) &  \sqrt{6}\sqrt{\lambda^2 + \kappa_j\lambda-6} [ (\kappa_j-\lambda)^2 - 6]\\
 - \sqrt{{3 \over 2}} (12 + \kappa_j^2 + \lambda \kappa_j)\sqrt{\lambda^2 + \kappa_j\lambda-6} &6 [ (\kappa_j - \lambda)^2 - 6]
\end{pmatrix} \, ,
\end{equation}
is exactly the stability matrix of the reduced two-field problem, while the other matrix is diagonal
\begin{equation}
 B_{\mathcal{N}-1 \times \mathcal{N}-1} =   \text{diag} \left({-3(\kappa_j -\kappa_i) \over \kappa_j +\lambda}, \cdots, 3+3w-2 \epsilon \right) \, ,
\end{equation}
for $i =1,\cdots,\mathcal{N}-2 \neq j$. 
The eigenvalues of $A$ need to satisfy the inequalities \eqref{eq:hyper1}-\eqref{eq:hyper2} (with the substitution $\kappa \rightarrow \kappa_j$) while the rest $\mathcal{N}-1$ eigenvalues are the diagonal elements of $B$ and so a stable solution requires
\begin{align} \label{eq:inequalities_kappa}
\kappa_j > \kappa_i~~\text{for}~~ \kappa_j>0\, , \qquad \kappa_j < \kappa_i~~\text{for}~~ \kappa_j< 0 \, ,
\end{align}
for $i\neq j$.
\begin{figure}[t!]
\centering
\includegraphics[width=.49\textwidth]{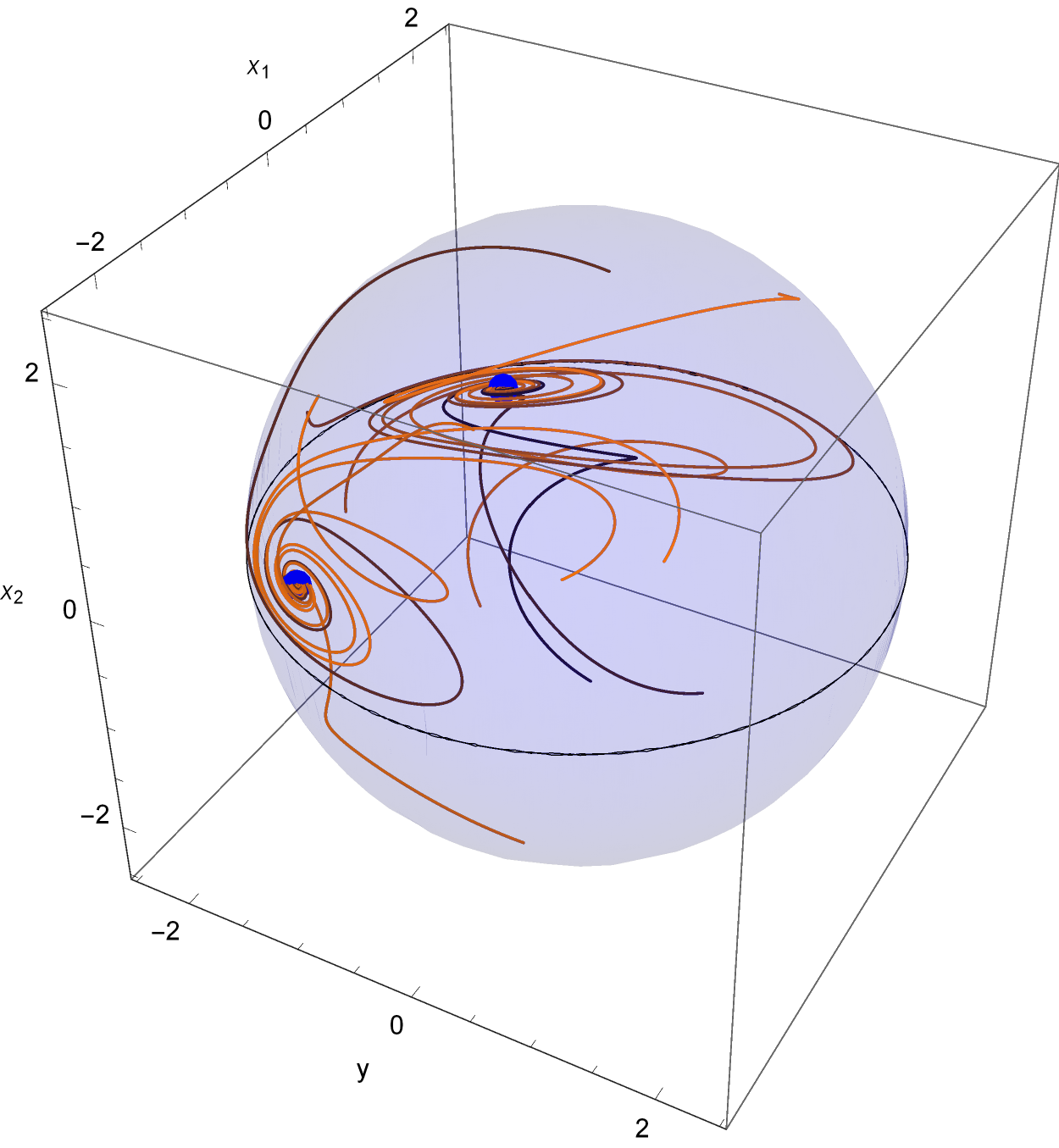}
\includegraphics[width=.49\textwidth]{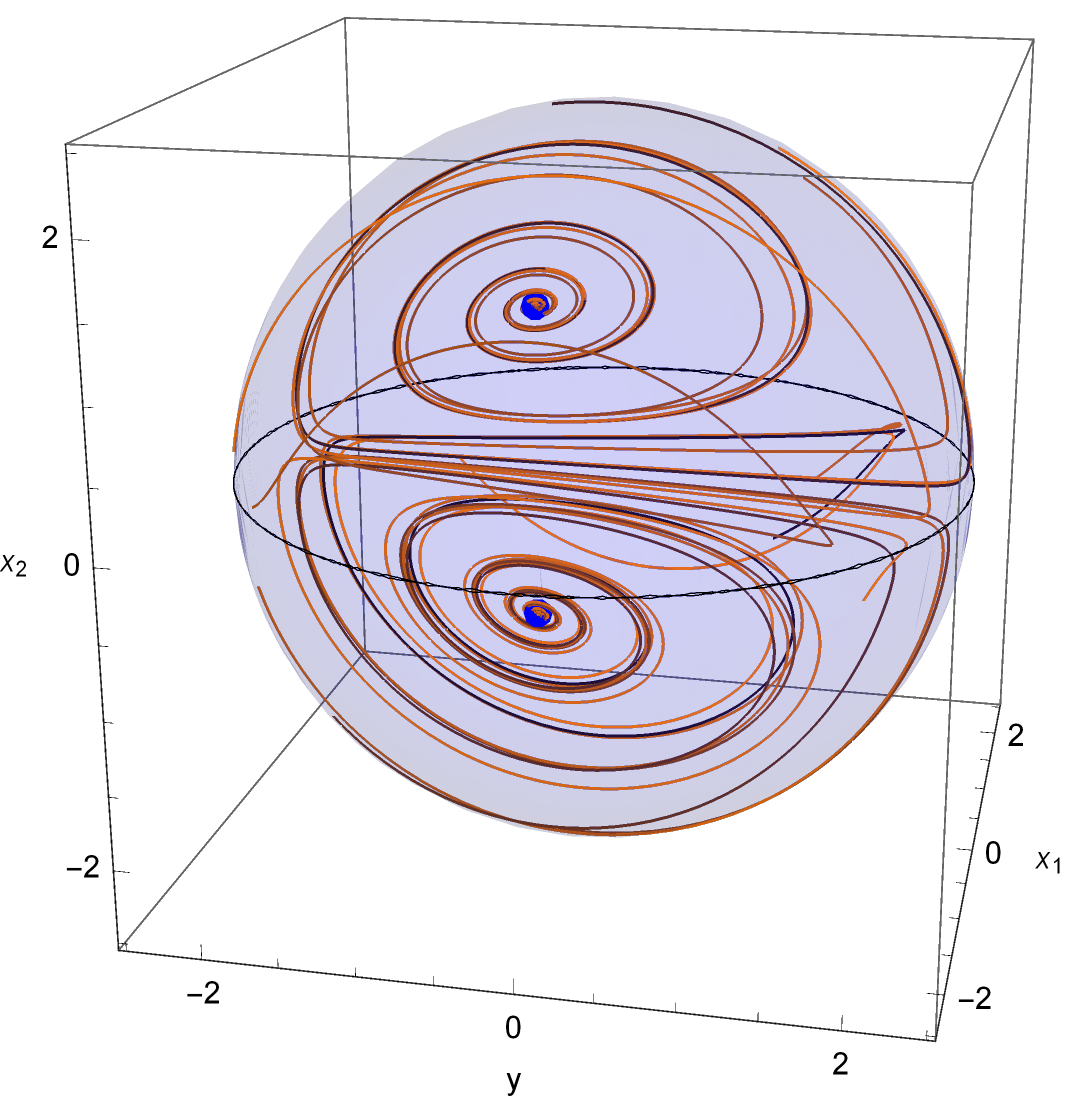}
\caption{ \it
The numerical solution for three fields and a wide range of initial conditions (drawn uniformly from the surface of sphere with $\epsilon\approx3$) with $\lambda=3$, $\kappa_1=1$, $z=0$ and $\kappa_2=-1$ (left) and $\kappa_2=10$ (right). The blue dots correspond to the two hyperbolic solutions. }
\label{fig:plot46}
\end{figure}

 \item The fluid domination and the scaling solution with non-zero $y,z$ and $x_i=0$
\begin{equation}
(y,x_i,z)_{\rm scal} = \left( -{3 (w+1)  \over \lambda},  0,\cdots ,3 - {9(1+w) \over \lambda^2} \right) \, ,
\end{equation}
both exist with the same stability properties as previously.

 \item Finally, in addition to the usual kinetic-domination solution
\begin{equation}
(y,x_i,z)_{\rm kin} = \left(\pm \sqrt{6},0,\cdots ,0 \right) \, ,
\end{equation}
which is stable for $|\lambda|>\sqrt{6}$ and $\lambda \kappa_i<0$, new solutions ($\epsilon=3$) exist with $y=0$ and $x_i\neq0$ which follow from
\begin{equation} \label{eq:kin}
\kappa_i x_i^2 = 0 \, , \qquad \sum x_i^2=6.
\end{equation}
For three fields the solution is trivially found to be
\begin{equation}
(y,x_1,x_2,z)_{\rm kin} = \left(0, \pm \sqrt{6}\sqrt{{ \kappa_1\over \kappa_1-\kappa_2}},\pm \sqrt{6}\sqrt{-{ \kappa_2 \over \kappa_1-\kappa_2}},0 \right) \, ,
\end{equation}
which exists provided $\kappa_1 \kappa_2<0$. It can be shown that the stability matrix evaluated on the solution contains at least one positive eigenvalue and thus this kinetic domination is unstable. For $\mathcal{N}>4$ fields this type of kinetic solution is defined on the $\mathcal{N}-2$ hypersurface containing points that satisfy the relations of \eqref{eq:kin} and it is again unstable.

\end{enumerate}
Having thoroughly examined the background in the next subsection we move to the study of quantum fluctuations neglecting the fluid's energy density and fluctuations. 

\subsection{Observables} \label{sec:observables}
To extract physical quantities it is necessary to express gauge-invariant perturbations in the orthonormal basis (local Frenet system) $Q^i = E^{~i}_A q^A$ where  the matrix $E^{~i}_{A}$ has as columns the components of the orthonormal vectors
\begin{equation}
E^{~i}_{A} = \left(t^i,n^i,b^i,\cdots \right) \, .
\end{equation}
Here, $t^i$ is the tangent unit vector, $n^i$ the normal vector, $b^i$ the binormal vector and so on. The projection along the tangent vector is related the curvature perturbation, while projections along the orthogonal directions are related to isocurvature perturbations. The covariant time derivatives of the orthonormal vectors satisfy the Frenet-Serret equations
\begin{equation}
\uD_t E^{~i}_{A} = C^{~B}_{A} E^{~i}_{B} \, ,
\end{equation}
where the matrix $C$ is antisymmetric with non-zero elements in the upper and lower diagonals.

It is well known that the curvature perturbation ($Q_{\sigma}\equiv Q_i E^{~i}_1 = q_1$) is sourced by the first isocurvature perturbation ($Q_{s}\equiv Q_i E^{~i}_2 = q_2$) \cite{Gordon:2000hv,Malik:2008im}, while the rest orthogonal perturbations interact via a ``mass matrix'' as well as through the curvatures that appear in the Frenet-Serret equations (see e.g.~Refs.~\cite{Kaiser:2012ak,Kaiser:2013sna,Aragam:2019omo} for examples including up to three fields and Refs.~\cite{Achucarro:2018ngj,Pinol:2020kvw} for a formal discussion including an arbitrary number of fields). More specifically, orthogonal fields are coupled through the following terms in the second order action
\begin{equation} \label{eq:matrix}
2\dot{q}^T \cdot \Omega \cdot q + q^T \cdot\Omega^T \cdot \Omega \cdot q- q^T \cdot M^2 \cdot q \, ,
\end{equation}
where $\Omega$ is the truncated matrix obtained from $C$ after removing elements of the first row and column, and the ``mass matrix'' is defined from
\begin{equation}
M^2_{AB}\equiv E^{~i}_{A} E^{~j}_{B} \left(V_{;ij} + \dot{\sigma}^2 t^k t^l R_{k i l j } + 3 \omega^2 \delta_{A2} \delta_{B2} \right) \, ,
\end{equation}
for $A,B>1$. Extra orthogonal fields decouple from $Q_{\sigma}$ and $Q_s$ when the matrices $\Omega$ and $M$ are block diagonal. A necessary condition for this decoupling is the vanishing of the torsion of the $\mathcal{N}$-dimensional field-space trajectory. Recall that the torsion of a curve is found by
calculating the rate of change of the normal unit vector. Using the
Frenet-Serret equations the torsion ($\tau$) is defined from
\begin{equation}
\uD_{t}n^{i} = -\omega t^{i}+\tau b^{i}\, ,
\end{equation}
where $t^{i},n^{i},b^{i}$ are the first three unit vectors of the orthonormal
frame at some point of the curve and $\omega$ is the turn rate. The case of zero torsion is reminiscent to geodesic motion where the curvature perturbation decouples from isocurvature perturbations.

For the hyperbolic solution the vectors $t^i,n^i$ have the first two components non-zero and the rest zero, which forces the next vectors in the series to have the first two components zero and the rest non-zero:
\begin{align}
t^i &= (t^{\phi},t^{\chi}, 0, \cdots) \, , \\
n^i &= (n^{\phi},n^{\chi}, 0, \cdots) \, , \\
b^i &= (0, 0, b^{\chi_2},\cdots) \, , \\
&\cdots
\end{align}
This means that the matrix $E^{~i}_{A}$ is block diagonal. Using the Frenet-Serret equation for $n^i$ we find $\text{D}_{t}n^{i}=-\omega t^{i}$ and, hence, the torsion is zero, which implies that $\Omega$ is block diagonal. Moreover, $M^2$ turns out to be block diagonal and so we conclude that $Q_{\sigma}$ and $Q_{s}$ evolve independently from the rest fields. Therefore, the basic predictions for this model, namely the spectral index and the tensor-to-scalar ratio, are identical to the two-field case.

\subsection{Field metric with $\mathcal{N}-1$ isometries}
Similar to Sec.~\ref{subsec:generic_2_field_metric} we can consider metrics that can support $\mathcal{N}-1$ integrals of motion, and hence they should have $\mathcal{N}-1$ isometries. This type of metrics naturally admit a $1 \times \mathcal{N}-1$ decomposition, with the $\phi$ field is canonically normalized and the rest $\mathcal{N}-1 \times \mathcal{N}-1$ matrix depending only on $\phi$. To simplify calculations and retain some analytical control we will consider the case of a diagonal metric 
\be
G_{ij} = \text{diag} (1, F_1(\phi),F_2(\phi),\cdots)\, ,
\ee
where the metric functions $F_1,F_2,\cdots$ can be different. Following the two-field discussion, in addition to the kinetic  and gradient solutions we can find de Sitter asymptotic solutions in the region where at least one of $\kappa_i \equiv (\ln F_i)_{, \phi}$ diverges and the inequalities \eqref{eq:inequalities_kappa} are satisfied.

\section{The general solution}

\label{sec:general_solution}
\subsection{Without matter source}

For the case without matter source we find that there are general solutions for the case where $\kappa$ and the lapse function are given by
\begin{equation} \label{eq:kappalambda}
\kappa=-\left(  \lambda+\sqrt{6}\right) \, , \qquad N_l \left(t\right)  =a^{-3-\sqrt{6}\lambda} \, .
\end{equation}
From the asymptotic analysis of Sec.~\ref{sec:two_fields} we know that this combination of $\kappa$ and $\lambda$ will give rise to either kinetic- ($|\lambda|>\sqrt{6}$) or gradient- ($|\lambda|<\sqrt{6}$) type solutions. Note also that for this choice of the lapse function $t$ is not the cosmic time, but rather another time variable. For the latter selection the field equations admit the Noetherian conservation laws%
\begin{equation}
I_{1}\left(  a,\dot{a},\phi,\dot{\phi},\psi,\dot{\psi}\right)  =\frac{\ud}%
{\ud t}\left(  a^{2 + \frac{\sqrt{6}}{2} \lambda }e^{-{1 \over 2} ( \lambda + \sqrt{6}) \phi}\right)  \, ,
\end{equation}%
\begin{equation}
I_{2}\left(  a,\dot{a},\phi,\dot{\phi},\psi,\dot{\psi}\right)  =a^{6 + \sqrt{6}\lambda}e^{-  (\lambda + \sqrt{6})\phi}\dot{\psi} \, .
\end{equation}
When $\lambda= - \sqrt{6}$ the field-space curvature is zero and the solution for this case can be found in Refs.~\cite{Chimento:1998ju,Christodoulidis:2018msl}. In the following we will consider $\lambda \neq -\sqrt{6}$.

Applying the coordinate transformation
\begin{equation} \label{eq:transformation}
\begin{aligned}
a\left(  \chi \left(  t\right)  ,\xi\left(  t\right)  ,\zeta \left(  t\right)  \right)
&  =a_{0}\left(  \left( \lambda + \sqrt{6} \right)^{2}\left(
\chi \left(  t\right)  \zeta \left(  t\right)  -\xi\left(  t\right)  ^{2}\right)
\right)  ^{\left(  6+\sqrt{6}\lambda\right)  ^{-1}} \, ,     \\
\phi\left(  \chi \left(  t\right)  ,\xi\left(  t\right)  ,\zeta \left(  t\right)
\right)   &  =\frac{2}{\lambda+\sqrt{6}}\ln\left(  \frac{\sqrt{  \left( \lambda + \sqrt{6} \right)^{2} \left(  \chi \left(  t\right)
\zeta \left(  t\right)  -\xi\left(  t\right)  ^{2}\right)   }}{2\chi \left(
t\right)  }\right) \, ,     \\
\psi\left(  \chi \left(  t\right)  ,\xi\left(  t\right)  ,\zeta \left(  t\right)
\right)   &  =\frac{\xi\left(  t\right)  }{\chi \left(  t\right)  } \, ,
\end{aligned}
\end{equation}
the point-like Lagrangian is expressed in the new coordinates as
\begin{equation}
L\left(  \chi ,\dot{\chi },\xi,\dot{\xi},\zeta ,\dot{\zeta }\right)  =2\left(  \dot{\xi}^{2}-\dot
{\chi }\dot{\zeta }\right)  -\bar{V}_{0}\chi ^{-\bar{\lambda}} \, , 
\end{equation}
where we defined for simplicity
\begin{equation}
\bar{V}_{0}=4^{-{\lambda \over \sqrt{6} + \lambda}} V_{0} \, , \qquad \bar{\lambda}= {2 \lambda \over \sqrt{6} + \lambda} \, .
\end{equation}
The field equations are written as follows%
\begin{equation}
\ddot{\xi}=0 \, , \qquad \ddot{\chi }=0 \,, \qquad \ddot{\zeta }-\frac{\bar{\lambda}\bar{V}_{0}}{2} \chi ^{\bar{\lambda}-1}=0 \, ,
\end{equation}
along with the constraint%
\begin{equation}
2\left(  \dot{\xi}^{2}-\dot{\chi }\dot{\zeta }\right)  +\bar{V}_{0}\chi ^{-\bar{\lambda}}=0 \, .
\end{equation}

Consequently, the analytic solution of the field equations is%
\[
\chi \left(  t\right)  =\chi _{0}\left(  t-t_{0}\right) \, , \qquad \xi\left(  t\right)
=\xi_{0}\left(  t-t_{1}\right) \, ,
\]
and%
\begin{equation}
\zeta \left(  t\right)  =\frac{\bar{V}_{0}\chi _{0}^{\bar{\lambda}-1}}{2\left(
\bar{\lambda}+1\right)  }\left(  t-t_{0}\right)  ^{\bar{\lambda}+1}%
+\zeta_{0}\left(  t-t_{2}\right) \, , \qquad \text{ for } \bar{\lambda}\neq-1,0 \, ,
\end{equation}
or%
\begin{equation}
\zeta \left(  t\right)  =\frac{\bar{V}_{0}}{2\chi _{0}^{2}}\ln\left(  t-t_{0}\right)
+\zeta_{0}\left(  t-t_{2}\right)  \, , \qquad \text{ for }\bar{\lambda}=-1 \, ,
\end{equation}
with the constraint $\xi_{0}^{2}-\chi _{0}\zeta_{0}=0$. For the special case of $\bar{\lambda}=0$, that is the potential is a cosmological constant with $\lambda=0$, the exact solution is%
\begin{equation}
\chi \left(  t\right)  =\chi _{0}\left(  t-t_{0}\right)  \, , \qquad \xi\left(  t\right)
=\xi_{0}\left(  t-t_{1}\right)  \, , \qquad \zeta \left(  t\right)  =\zeta_{0}\left(
t-t_{2}\right) \, ,
\end{equation}
with the constraint equation $\xi_{0}^{2}-\chi _{0}\zeta_{0}=0$. With $\chi,\xi$ and $\zeta$ known the solution in terms of the original variables of the problem can be found using the transformation \eqref{eq:transformation}.

\begin{figure}[ptb]
\centering\includegraphics[width=0.45\textwidth]{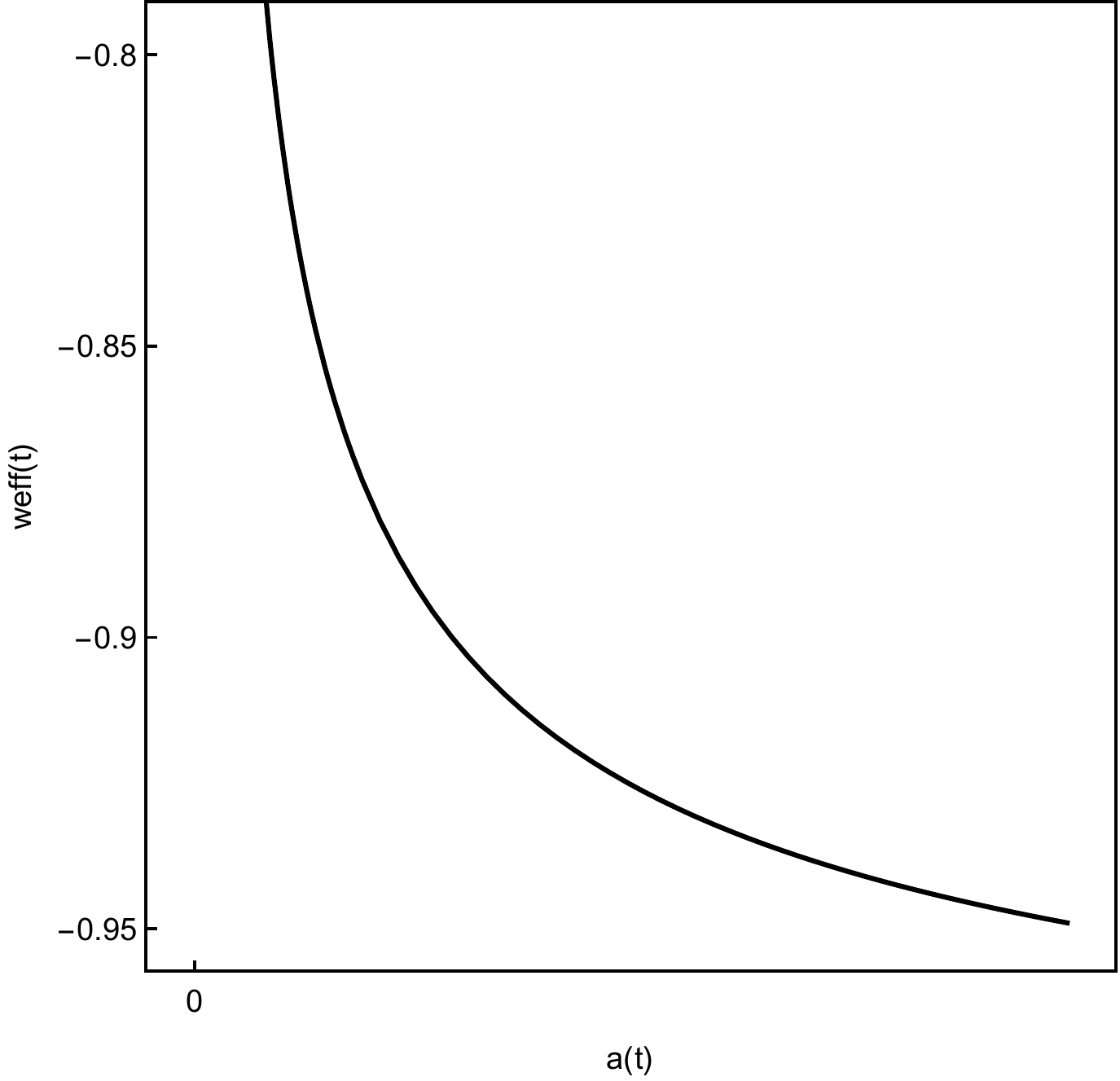}
\centering\includegraphics[width=0.45\textwidth]{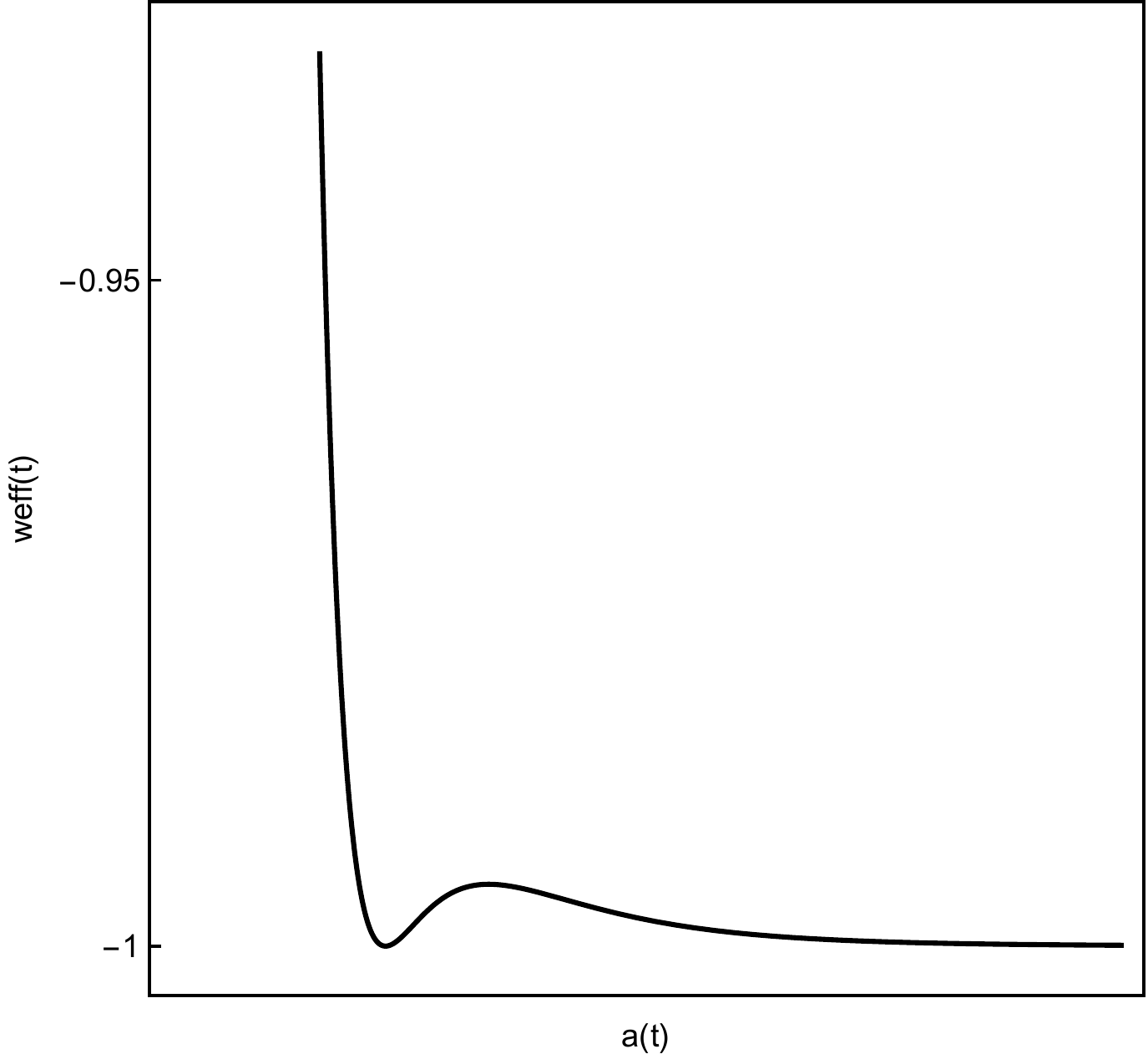}
\caption{Qualitative evolution of the effective equation of state parameter
$w_{\rm eff}=-1-\frac{2}{3}\frac{\dot{H}}{H^{2}}$ for the analytic solution of our
consideration for $\left(  \lambda,\chi _{0},\zeta_{0},t_{0},t_{1},t_{2},\bar{V}%
_{0}\right)$ equal to  $\left(  -1,0.1,0.1,0,0.1,10,1\right)  $ (left) and $\left(  \sqrt{6},1,\zeta_{0},0,0,0,1\right)  $ (right).}
\label{fig1A}%
\end{figure}

In Fig.~\ref{fig1A} we plot the qualitative evolution of
the effective equation of state parameter $w_{\rm eff}=w_{\rm eff}\left(  a\right)  $,
which is defined as $w_{\rm eff}\left(  a\left(  t\right)  \right)  =-1-\frac
{2}{3}\frac{\dot{H}}{H^{2}}$, for two different values of the parameter
$\lambda$. We observe that the $w_{\rm eff}\left(  a\right)  $ is that of the hyperbolic expansion.

It was pointed out recently in Ref.~\cite{anlp} that when the second field is phantom we can reconstruct the analytic solution for the two-field model in a same way as before by setting $\tilde{\psi} =i \psi$. Therefore, the analytic solution for the second case for $\kappa=\left(  \lambda+\sqrt {6}\right)  $ is determined in a similar way and we omit the presentation.
 Indeed by considering $\xi(t)=i \xi(t)$ we obtain the coordinate transformation

\begin{align}
a\left(  \chi \left(  t\right)  ,\xi\left(  t\right)  ,\zeta \left(  t\right)  \right)
&  =a_{0}\left(   \left( \lambda + \sqrt{6} \right)^{2} \left(
\chi \left(  t\right) \zeta \left(  t\right)  +\xi\left(  t\right)  ^{2}\right)
\right)  ^{\left(  6+\sqrt{6}\lambda\right)  ^{-1}} \, , \\
\phi\left(  \chi \left(  t\right)  ,\xi\left(  t\right)  ,\zeta  \left(  t\right)
\right)   &  =\frac{2}{\lambda+\sqrt{6}}\ln\left(  \frac{\sqrt{\left(
 \left( \lambda + \sqrt{6} \right)^{2}  \left(  \chi \left(  t\right)
\zeta \left(  t\right)  +\xi\left(  t\right)  ^{2}\right)  \right)  }}{2\chi \left(
t\right)  }\right)  \, ,  \\
\psi\left(  \chi \left(  t\right)  ,\xi\left(  t\right)  ,\zeta  \left(  t\right)
\right)   &  =\frac{\xi\left(  t\right)  }{\chi \left(  t\right)  } \, ,
\end{align}
which produces the same second-order differential equations as before.
Consequently, the solution for the variables $\chi \left(  t\right)  ,~\xi\left(
t\right)  $ and $\zeta  \left(  t\right)  $ is the same as before, while the
constraint equation is now $\xi_{0}^{2}+\chi _{0}\zeta_{0}=0$.

\subsection{In the presence of matter source}

Consider now the existence of an additional matter field. We conclude that
when the equation of state parameter is $w\left(  \lambda\right)  =-1+\frac
{\sqrt{6}}{3}\lambda$, then the solutions found in the previous section hold for this cosmological model as well, with the only difference that the constraint equations for the integration constants $\xi_{0},\chi_{0},\zeta_{0}$ are now
\begin{equation}
\rho_{m0}=2\xi_{0}^{2}-\chi_{0}\zeta_{0} \, ,
\end{equation}
for the Chiral model and
\begin{equation}
\rho_{m0}=\xi_{0}^{2}+\chi_{0}\zeta_{0} \, ,
\end{equation}
for the Chiral-quintom model.

We observe that for $w_{m}=-\frac{1}{3}$ the point-like Lagrangian (\ref{ssl1})
describes the gravitational field equations for the case of a FLRW spacetime
with spatially curvature $k=\rho_{m0}$, that is, for the line element%
\begin{equation}
\ud s^{2}=-N^{2}\left(  t\right)  \ud t^{2}+\frac{a^{2}\left(  t\right)  }%
{1-\frac{k}{4}\left(  X^{2}+Y^{2}+ Z^{2}\right)  }\left(  \ud X^{2}+ \ud Y^{2}%
+\ud Z^{2}\right)  \, .
\end{equation}
Hence, we also presented for the first time an analytic solution for a two-field
model in a non-flat FLRW background space.

\section{Discussion}

\label{sec:conclusions}

In this work we studied the dynamics and the existence of analytic solutions
for a multi-field cosmological model in a spatially flat FLRW background
space. In particular, we considered a cosmological model consisting of
$\mathcal{N}$ scalar fields minimally coupled to gravity with a hyperbolic
interaction in the kinetic terms in the presence of an ideal gas. For
this gravitational model we studied 
the asymptotic behaviour of the field equations. We recovered
previous results for the two-field model, that is all critical points for the quintessence, i.e. of
single-field theory, as well as  a pair of points which describe a
hyperbolic solution where both fields contribute.

In the multi-field scenario with $\mathcal{N}>2$ fields we showed that the number of dynamical fields for any late-time solution, and most importantly for the hyperbolic one, remains two. Therefore, at the background level the late-time behaviour of this problem is identical to the two-field case. Moreover, in Sec.~\ref{sec:observables} we showed that the same is true for first order perturbations. This is an important observation because it is clear that by adding additional scalar fields in this specific theory we do not get new physical results (regarding late-time solutions), while no information can be extracted from current observations to support this $\mathcal{N}$-field theory with $\mathcal{N}>2$. It would be interesting to investigate whether this holds for non-Gaussianities and other higher-order correlators as well.

Finally, for the two-field model with an exponential potential we
proved for the first time the Liouville integrability of the field equations
while we derive the analytic solution of the model. We applied the theory of
similarity transformations to construct conservation laws. We found that for a
specific combination of the exponents $\lambda$ and $\kappa$ (associated with the potential gradient and the curvature of the hyperbolic space respectively) new conservation laws
exist which facilitate the derivation of a closed-form solution of the dynamical
system. We demonstrated that the solution holds for the case of the presence of
additional matter, while it can be used to construct the analytic solution of
the multi-field model in a non-spatially flat FLRW background space.

\begin{acknowledgments}
PC acknowledges financial support from the Dutch Organisation for Scientific Research (NWO).
\end{acknowledgments}


\begin{thebibliography}{99}



\bibitem{Brown:2017osf}
A.~R.~Brown,
``Hyperbolic Inflation,''
Phys. Rev. Lett. \textbf{121} (2018) no.25, 251601
[arXiv:1705.03023 [hep-th]].

\bibitem{Mizuno:2017idt}
S.~Mizuno and S.~Mukohyama,
``Primordial perturbations from inflation with a hyperbolic field-space,''
Phys. Rev. D \textbf{96} (2017) no.10, 103533
[arXiv:1707.05125 [hep-th]].

\bibitem{Cicoli:2020cfj}
M.~Cicoli, G.~Dibitetto and F.~G.~Pedro,
``New accelerating solutions in late-time cosmology,''
Phys. Rev. D \textbf{101} (2020) no.10, 103524
[arXiv:2002.02695 [gr-qc]].

\bibitem{Cicoli:2020noz}
M.~Cicoli, G.~Dibitetto and F.~G.~Pedro,
``Out of the Swampland with Multifield Quintessence?,''
JHEP \textbf{10} (2020), 035
[arXiv:2007.11011 [hep-th]].

\bibitem {Fumagalli:2019noh}J.~Fumagalli, S.~Garcia-Saenz, L.~Pinol,
S.~Renaux-Petel and J.~Ronayne, ``Hyper-Non-Gaussianities in Inflation with
Strongly Nongeodesic Motion,'' Phys. Rev. Lett. \textbf{123} (2019) no.20,
201302
[arXiv:1902.03221 [hep-th]].

\bibitem {Bjorkmo:2019qno}T.~Bjorkmo, R.~Z.~Ferreira and M.~C.~D.~Marsh,
``Mild Non-Gaussianities under Perturbative Control from Rapid-Turn Inflation
Models,'' JCAP \textbf{12} (2019), 036
[arXiv:1908.11316 [hep-th]].

\bibitem{Ferreira:2020qkf}
R.~Z.~Ferreira,
``Non-Gaussianities in models of inflation with large and negative entropic masses,''
JCAP \textbf{08} (2020), 034
[arXiv:2003.13410 [astro-ph.CO]].


\bibitem{Bounakis:2020xaw}
M.~Bounakis, I.~G.~Moss and G.~Rigopoulos,
``Observational constraints on Hyperinflation,''
[arXiv:2010.06461 [gr-qc]].

\bibitem{Basilakos:2019dof}
S.~Basilakos, G.~Leon, G.~Papagiannopoulos and E.~N.~Saridakis,
``Dynamical system analysis at background and perturbation levels: Quintessence in severe disadvantage comparing to $\Lambda$CDM,''
Phys. Rev. D \textbf{100} (2019) no.4, 043524
[arXiv:1904.01563 [gr-qc]].

\bibitem{Banerjee:2020xcn}
A.~Banerjee, H.~Cai, L.~Heisenberg, E.~\'O.~Colg\'ain, M.~M.~Sheikh-Jabbari and T.~Yang,
[arXiv:2006.00244 [astro-ph.CO]].
 
\bibitem{Bjorkmo:2019aev}
T.~Bjorkmo and M.~C.~D.~Marsh,
``Hyperinflation generalised: from its attractor mechanism to its tension with the \textquoteleft{}swampland conditions\textquoteright{},''
JHEP \textbf{04} (2019), 172
[arXiv:1901.08603 [hep-th]].


\bibitem{Aragam:2020uqi}
V.~Aragam, S.~Paban and R.~Rosati,
``The Multi-Field, Rapid-Turn Inflationary Solution,''
[arXiv:2010.15933 [hep-th]].

\bibitem {Chimento:1998ju}L.~P.~Chimento, ``General solution to two-scalar
field cosmologies with exponential potentials,'' Class. Quant. Grav.
\textbf{15} (1998), 965-974


\bibitem {Christodoulidis:2018msl}P.~Christodoulidis, ``Probing the
inflationary evolution using analytical solutions,'' [arXiv:1811.06456 [astro-ph.CO]].

\bibitem {Paliathanasis:2018vru}A.~Paliathanasis, G.~Leon and S.~Pan, ``Exact
Solutions in Chiral Cosmology,'' Gen. Rel. Grav. \textbf{51} (2019) no.9, 106
[arXiv:1811.10038 [gr-qc]].



\bibitem {Socorro:2020nsm}J.~Socorro, S.~P\'erez-Pay\'an, R.~Hern\'andez,
A.~Espinoza-Garc\'\i{}a and L.~R.~D\'\i{}az-Barr\'on,
``Classical and quantum exact solutions for a FRW in chiral like cosmology,''
[arXiv:2012.11108 [gr-qc]].


\bibitem{Socorro:2021bco}
J.~Socorro, S.~P\'erez-Pay\'an, A.~Espinoza-Garc\'\i{}a and L.~R.~D\'\i{}az-Barr\'on,
[arXiv:2101.05973 [gr-qc]].


\bibitem {ns1}S.~Basilakos, M.~Tsamparlis and A.~Paliathanasis,
 ``Using the Noether symmetry approach to probe the nature of dark energy,''
Phys. Rev. D \textbf{83} (2011), 103512 
[arXiv:1104.2980 [astro-ph.CO]].

 


\bibitem {ns2}A.~Paliathanasis, M.~Tsamparlis and S.~Basilakos,
``Constraints and analytical solutions of $f(R)$ theories of gravity using Noether symmetries,''
Phys. Rev. D \textbf{84} (2011), 123514 
[arXiv:1111.4547 [astro-ph.CO]].



\bibitem {ns3}J.~A.~Belinch\'on, T.~Harko and M.~K.~Mak,
Astrophys. Space Sci. \textbf{361} (2016) no.2, 52
doi:10.1007/s10509-015-2642-7 [arXiv:1512.08054 [gr-qc]].



\bibitem {ns4}M.~Demianski, E.~Piedipalumbo, C.~Rubano and C.~Tortora,
``Accelerating universe in scalar tensor models: Confrontation of theoretical predictions with observations,''
Astron. Astrophys. \textbf{454} (2006), 55-66 
[arXiv:astro-ph/0604026 [astro-ph]].



\bibitem {ns5}P.~A.~Terzis, N.~Dimakis and T.~Christodoulakis,
``Noether analysis of Scalar-Tensor Cosmology,''
Phys. Rev. D \textbf{90} (2014) no.12, 123543
[arXiv:1410.0802 [gr-qc]].



\bibitem {ns6}A.~Paliathanasis, J.~D.~Barrow and P.~G.~L.~Leach,
``Cosmological Solutions of $f(T)$ Gravity,''
Phys. Rev. D \textbf{94} (2016) no.2, 023525
[arXiv:1606.00659 [gr-qc]].



\bibitem {ns7}J.~D.~Barrow, S.~Cotsakis and A.~Tsokaros,
``Series expansions and sudden singularities,''
d
[arXiv:1301.6523 [gr-qc]].



\bibitem {ns8}S.~Cotsakis,
``Asymptotic Poincar\'e compactification and finite-time singularities,''
Grav. Cosmol. \textbf{19} (2013), 240-245 
[arXiv:1301.4778 [gr-qc]].



\bibitem {dn1}L.~Amendola, R.~Gannouji, D.~Polarski and S.~Tsujikawa,
``Conditions for the cosmological viability of f(R) dark energy models,''
Phys. Rev. D \textbf{75} (2007), 083504
[arXiv:gr-qc/0612180 [gr-qc]].




\bibitem {dn2}A.~Coley and G.~Leon,
``Static Spherically Symmetric Einstein-aether models I: Perfect fluids with a linear equation of state and scalar fields with an exponential self-interacting potential,''
Gen. Rel. Grav. \textbf{51} (2019) no.9, 115 
[arXiv:1905.02003 [gr-qc]].




\bibitem {dn3}G.~Leon, Y.~Leyva and J.~Socorro,
``Quintom phase-space: beyond the exponential potential,''
Phys. Lett. B \textbf{732} (2014), 285-297
[arXiv:1208.0061 [gr-qc]].




\bibitem {dn4}R.~Lazkoz, G.~Leon and I.~Quiros,
``Quintom cosmologies with arbitrary potentials,''
Phys. Lett. B \textbf{649} (2007), 103-110
[arXiv:astro-ph/0701353 [astro-ph]].


\bibitem {quin00}Y.F. Cai, E.N. Saridakis, M.R.\ Setare and J.-Q. Xia, ``Quintom cosmology: theoretical implications and observations,'' Phys.
Rep. 493, 1 (2010)


\bibitem {chir3}S.V. Chervon, ``On the chiral model of cosmological inflation,'' Russ. Phys. J. 38, 539 (1995)

\bibitem {sigm0}S. V. Ketov, ``Quantum Non-linear Sigma Models,'', Springer-Verlag,
Berlin, (2000).

\bibitem {sigm1}J. Lee, T.H. Lee, T. Moon and P. Oh, ``de Sitter nonlinear sigma model and accelerating universe,'' Phys. Rev. D 80, 065016 (2009)

\bibitem {ch1}S.V. Chervon, ``Chiral Cosmological Models: Dark Sector Fields Description,'' Quantum Matter 2, 71 (2013)

\bibitem {ancqg}A. Paliathanasis, ``
Dynamics of chiral cosmology'', Class. Quantum Grav. 37, 195014 (2020)

\bibitem {Christodoulidis:2019jsx}P.~Christodoulidis, D.~Roest and
E.~I.~Sfakianakis, ``Scaling attractors in multi-field inflation,'' JCAP
\textbf{12}, 059 (2019)
[arXiv:1903.06116 [hep-th]].

\bibitem {Dimakis:2019qfs}N.~Dimakis, A.~Paliathanasis, P.~A.~Terzis and
T.~Christodoulakis, Eur. Phys. J. C \textbf{79}, no.7, 618 (2019)
[arXiv:1904.09713 [gr-qc]].

\bibitem {Paliathanasis:2020abu}A.~Paliathanasis and G.~Leon, ``Asymptotic behavior of N-fields Chiral Cosmology.'' Eur. Phys. J. C
\textbf{80} (2020) no.9, 847
[arXiv:2007.13223 [gr-qc]].

\bibitem{Gordon:2000hv}
C.~Gordon, D.~Wands, B.~A.~Bassett and R.~Maartens,
``Adiabatic and entropy perturbations from inflation,''
Phys. Rev. D \textbf{63} (2000), 023506
[arXiv:astro-ph/0009131 [astro-ph]].

\bibitem{Malik:2008im}
K.~A.~Malik and D.~Wands,
``Cosmological perturbations,''
Phys. Rept. \textbf{475} (2009), 1-51
[arXiv:0809.4944 [astro-ph]].

\bibitem{Kaiser:2012ak}
D.~I.~Kaiser, E.~A.~Mazenc and E.~I.~Sfakianakis,
``Primordial Bispectrum from Multifield Inflation with Nonminimal Couplings,''
Phys. Rev. D \textbf{87} (2013), 064004
[arXiv:1210.7487 [astro-ph.CO]].


\bibitem{Kaiser:2013sna}
D.~I.~Kaiser and E.~I.~Sfakianakis,
``Multifield Inflation after Planck: The Case for Nonminimal Couplings,''
Phys. Rev. Lett. \textbf{112} (2014) no.1, 011302
[arXiv:1304.0363 [astro-ph.CO]].

 
\bibitem{Aragam:2019omo}
V.~Aragam, S.~Paban and R.~Rosati,
``Multi-field Inflation in High-Slope Potentials,''
JCAP \textbf{04} (2020), 022
[arXiv:1905.07495 [hep-th]].

\bibitem{Achucarro:2018ngj}
A.~Ach\'ucarro, S.~C\'espedes, A.~C.~Davis and G.~A.~Palma,
``Constraints on Holographic Multifield Inflation and Models Based on the Hamilton-Jacobi Formalism,''
Phys. Rev. Lett. \textbf{122} (2019) no.19, 191301
[arXiv:1809.05341 [hep-th]].

\bibitem{Pinol:2020kvw}
L.~Pinol,
``Multifield inflation beyond $N_\mathrm{field}=2$: non-Gaussianities and single-field effective theory,''
[arXiv:2011.05930 [astro-ph.CO]].


\bibitem {anlp}A. Paliathanasis and G. Leon, Dynamics of a two scalar field
cosmological model with phantom terms, [arXiv:2009.12874]

\end{thebibliography}
\end{document}